\documentclass[journal]{IEEEtran}
\usepackage{graphicx}

\begin{document}
\title{Geant4-related R\&D for new particle transport methods}

\author{M. Augelli, M. Begalli, T. Evans, E. Gargioni, S. Hauf, C. H. Kim,
        M. Kuster, M. G. Pia, P. Queiroz Filho, L. Quintieri, P. Saracco, 
        D. Souza Santos,  G. Weidenspointner, A. Zoglauer% <-this % stops a space
\thanks{Manuscript received November 20, 2009.} % <-this % stops a space
\thanks{Mauro Augelli is with the Centre d'Etudes Spatiales (CNES), Toulouse,
	France.}
\thanks{Marcia Begalli is with State University of Rio de Janeiro, Brazil.}
\thanks{Thomas Evans is with Oak Ridge National Laboratory, Oak Ridge, TN, USA.}
\thanks{Elisabetta Gargioni is with University Medical Center 
	Hamburg-Eppendorf, Germany.}
\thanks{Steffen Hauf and Markus Kuster are with University of Technology,
	Darmstadt, Germany.}
\thanks{Chan Hyeong Kim is with Hanyang University, Seoul, Korea.}
\thanks{Maria Grazia Pia and Paolo Saracco are with INFN Sezione di Genova, 
	Via Dodecaneso 33, 16146 Genova, Italy (e-mail:
	MariaGrazia.Pia@ge.infn.it, Paolo.Saracco@ge.infn.it).}
\thanks{Pedro Queiroz Filho and Denison Souza Santos are with Institute 
	for Radiation Protection and Dosimetry IRD, Rio de Janeiro, Brazil.}
\thanks{Lina Quintieri is with INFN Laboratori Nazionali di Frascati, 
	Frascati, Italy.}
%\thanks{Manju Sudhakar is with INFN Sezione di Genova, Via
%	Dodecaneso 33, 16146 Genova, Italy (e-mail:Manju.Sudhakar@ge.infn.it); 
%	she is on leave from ISRO, Bangalore, India and Phys. Dept.,
%        Univ. of Calicut, India.}
\thanks{Georg Weidenspointner is with the Max-Planck-Institut f\"ur
	extraterrestrische Physik, Postfach 1603, 85740 Garching, Germany, and
	with the MPI Halbleiterlabor.}
\thanks{Andreas Zoglauer is with the Space Sciences Laboratory,
	University of California at Berkeley, Berkley, CA, USA.}%
}

\maketitle
\pagestyle{empty}
\thispagestyle{empty}

\begin{abstract}
A R\&D project has been launched in
2009 to address fundamental methods in radiation transport simulation
and revisit Geant4 kernel design to cope with new experimental
requirements. The project focuses on simulation at different scales in
the same experimental environment: this set of problems requires new
methods across the current boundaries of condensed-random-walk and
discrete transport schemes. An exploration is also foreseen about
exploiting and extending already existing Geant4 features to apply
Monte Carlo and deterministic transport methods in the same simulation
environment.  An overview of this new R\&D associated with Geant4 is
presented, together with the first developments in progress.

\end{abstract}

%\begin{IEEEkeywords}
%IEEEtran, journal, \LaTeX, paper, template.
%\end{IEEEkeywords}

\section{Introduction}
\IEEEPARstart{G}{eant4} \cite{g4nim,g4tns} is an object oriented toolkit for the
simulation of particle interactions with matter. It provides advanced
functionality for all the domains typical of detector simulation:
geometry and material modelling, description of particle properties,
physics processes, tracking, event and run management, user interface
and visualisation.

Geant4 is nowadays a mature Monte Carlo system; its multi-disciplinary
nature and its wide usage are demonstrated by the fact that its
reference article \cite{g4nim} is the most cited publication 
\cite{swpub,g4cite} in the
Nuclear Science and Technology category.

Geant4 is the result of a R\&D project (CERN RD44) \cite{rd44} 
carried out between
1994 and 1998. RD44 was launched at a time when the LEP experiments
were running GEANT 3 as a well-established system, that had been
refined throughout a decade of production service. RD44 investigated
the adoption of the object oriented technology and C++ for a
simulation system replacing GEANT 3.21 \cite{geant3}, and developed the first
functional version of the Geant4 released at the end of 1998.

The Geant4 releases following 
the first Geant4 one at the end of 1998 have added new functionality to
the toolkit; nevertheless, its architectural
design and the fundamental concepts of Geant4 application domain
have remained substantially unchanged since their original conception.

New experimental requirements have emerged in the recent years, which
challenge the conventional scope of major Monte Carlo transport codes
like Geant4. Research in nanodosimetry, nanotechnology-based
detectors, radiation effects on components in space and at high
luminosity colliders, nuclear power, plasma physics etc. have
shown the need of new methodological approaches to radiation transport
simulation along with new physics functionality in Geant4.
A common requirement has emerged in all such research domains,
i.e. the ability to change the scale at which the problem is treated
in a complex simulation environment. This requirement goes
beyond the traditional issues of variance reduction, for which current
Monte Carlo codes provide abundant tools and techniques.

Significant technological developments both in software and
computing hardware have also occurred since the RD44 phase. New
software techniques are available nowadays, that were not yet
established at the time when Geant4 was designed.

\section{Recent experimental evolutions and requirements}

Various experimental domains would profit of the
capability of dealing with co-working transport schemes. 

Radiation effects at the nano-scale are
important for the protection of electronic devices operating in
various radiation environments. In realistic use cases 
small-scale systems are often embedded in larger scale ones: for
instance, a component may operate within a collider experiment or on a
satellite in space, cellular and sub-cellular aggregates in real
biological systems exist in complex body structures etc.
A simulation system capable of addressing different scales would be
relevant to studies of the effects on components exposed to the fierce
experimental environment of LHC (and future super-LHC).

The capability of condensed and discrete Monte Carlo simulation
schemes in the same software environment is also critical to
experimental configurations involving nanotechnology-based
detectors. While R\&D in nanotechnology is actively pursued also in
view of application to HEP detectors, it is not yet possible to
simulate such detectors as standalone systems with Geant4, nor to
evaluate their performance once they are embedded in a full-scale
experimental set-up.

Plasma physics requires addressing the concept of object state and
behaviour mutation in relation to the environment: in this use case
the mutability concerns both the physics processes and the particles
involved. Astrophysics and studies for fusion-based nuclear power are
just two relevant applications, which would profit of Geant4
applicability to this physics domain.

The issue of co-existing condensed-random-walk and discrete schemes
arises in another context of the simulation domain. It concerns the
conceptually correct treatment of the atomic relaxation following the
impact ionization produced by charged particles: since the cross
section for producing secondary electrons from ionization is subject
to infrared divergence, in the conventional condensed-random-walk
schemes the interaction is treated in two different regimes of
continuous energy loss along the step and of discrete X-ray
production, with the consequent adoption of cutoffs. This scheme
introduces an artificial dependency on cuts in the generation of PIXE
(Particle Induced X-ray Emission), while atomic relaxation is
intrinsically a discrete process; moreover, the current Geant4 scheme
neglects the correlation between the X-ray spectrum of primary
ionization and PIXE. A conceptual revision of the continuous
energy loss and discrete scheme would be desirable in this physics domain
too.
Use cases affected by the current conceptual limitation to treat PIXE
correctly in Geant4 involve multiple, multi-disciplinary domains:
applications for material analysis from planetary science to cultural
heritage, precise dosimetry, critical shielding optimization of X-ray
telescopes etc.

\section{Main areas of research}

A R\&D project, named NANO5,
 has been recently launched \cite{mc2009} to address fundamental
methods in radiation transport simulation; it explores possible
solutions to cope with the new experimental requirements mentioned in
the previous section and evaluates whether and how they can be
supported by Geant4 kernel design.
It was initiated by a team at the Italian Institute of Nuclear
Research (INFN) and currently gathers an international team of
physicists and engineers with background in various disciplines.

The main focus of the project lies in the simulation at different
scales in the same experimental environment: this objective is
associated with the research of transport methods across the current
boundaries of condensed-random-walk and discrete transport schemes.
An exploration is also foreseen about exploiting and extending already
existing Geant4 features to apply Monte Carlo and deterministic
transport methods in the same simulation environment.
The Geant4 toolkit is the ideal playground for this research, thanks
to the object oriented technology it adopted in the RD44 phase.

Other issues have been identified along with the experience of Geant4
development and usage over the past 10 years, which would profit of
the exploratory research in the kernel design motivated by the main objectives
of the project:
\begin{itemize}
\item Customization of physics modeling in a simulation application
\item Scattered and tangled concerns across the code
\item Facilities for physics verification and validation
\item Performance
\end{itemize}
These topics are considered as supporting developments, which are
instrumental to achieve the main goals of the project.

The project adopts a software process
model based on the Unified Process \cite{up} framework.
The software developments are motivated by concrete
experimental applications, and significant effort is invested in the
software design: these features of the project are well served by the
Unified Process, which is use case driven and architecture-centric.
The adopted software process framework involves an iterative and
incremental life-cycle.

\section{Co-working condensed and discrete simulation methods}

Methods to model hard interactions of particles with matter
constituents by means of an appropriate binary theory are well
established: in this approach collisions are treated as binary
processes, that is, either the target electrons are treated as free
and at rest, or the influence of binding is accounted only in an
approximated way.

General-purpose Monte Carlo codes, like EGS \cite{egs4,egsnrc,egs5}, 
FLUKA \cite{fluka1,fluka2},
Geant4 and MCNP \cite{mcnp,mcnp5,mcnpx}, operate in this context. 
Their calculations
of energy deposit distributions are based on condensed-random-walk
schemes of particle transport. Charged particle tracks are divided
into many steps, such that several interactions occur in a step; one
energy loss and one deflection are calculated for each step. A further
simplification consists in the adoption of the Continuous Slowing Down
Approximation (CSDA), where the energy loss rate is determined by the
stopping power. This approach is adequate as long as the discrete
energy loss events treated are of magnitudes larger than electronic
binding energies.

Various specialized Monte Carlo codes, usually known as ``track
structure codes'', have been developed for micro and nano-dosimetry
calculations. They handle particle interactions with matter as
discrete processes: all collisions are explicitly simulated as
single-scattering interactions. This approach is suitable to studies
where the precise structure of the energy deposit and of the secondary
particle production associated with a track is essential;
nevertheless, the detailed treatment of collisions down to very low
energy results in a high computational demand.

So far, simulation based on condensed-random-walk schemes and so-called track
structure generation have been treated as distinct computational
domains; this separation is due to the conceptual and technical
difficulty of handling the two schemes in the same simulation
environment. Achieving a conceptual approach and an architectural
design where the two schemes can co-work would represent a significant
progress in Monte Carlo simulation.

Recently, a set of specialized processes for track structure
simulation in liquid water has been designed and implemented in Geant4
\cite{dna}; like their equivalents in dedicated Monte Carlo codes, they
operate in the regime of discrete interactions. While the toolkit
nature of Geant4 allows the co-existence of tools for simulation at
different scales, the capability of these two schemes to effectively
co-work in a multi-scale problem is still far from being established.

\section{Co-working Monte Carlo and deterministic simulation methods}

Deterministic transport methods are widely used in various domains:
radiotherapy treatment planning and calculations for nuclear reactors
are just two examples. Their usage is motivated by the requirement of
a fast simulation response in complex situations.

Both Monte Carlo and deterministic transport methods have their own
strong points and limitations. Research in this field to perform both
transport models within the same simulation environment would be
highly valuable: the capability of different transport methods in the
same simulation environment would simplify the geometrical and
material modelling of the system under study, and would facilitate the
analysis of the behaviour of the system itself.

\section{Supporting research topics}
The complexity of the problem domain to be addressed requires the investigation of software techniques, capable to effectively support the conceptual objectives to be pursued.

\subsection{Generic programming techniques in physics simulation design}

Metaprogramming has emerged in the last few years as a powerful software
technique. In C++ the template mechanism provides naturally a rich
facility for metaprogramming; libraries like Boost \cite{boost}
and Loki \cite{alexandrescu} are
nowadays available to support generic programming
development. Metaprogramming presents several interesting advantages,
which propose it as a worthy candidate for physics simulation design.

This technique has not been exploited in Geant4 core yet: the
evolution towards the C++ standard still in progress and the limited
support available in C++ compilers at the time of the RD44 phase discouraged 
the consideration of templates for extensive use in Geant4 architectural 
design at that stage. 

A preliminary investigation of its applicability in a
multi-platform simulation context has been carried out by one of the
authors of this paper through the application of a policy-based class
design \cite{dna} limited to a restricted physics sub-domain.

\subsection{Design for scattered and tangled concerns}

The problem domain of radiation transport simulation involves a number
of concerns, which are common to multiple parts of the system, but
whose code gets scattered across different parts; moreover, multiple
concerns may be tangled in the same code. The capability of addressing
scattered concerns by an effective design would results in leaner,
more easily maintainable code: an optimization in this domain would be
meaningful in a large-scale software system like Geant4.

The object oriented technology lacks proper instruments to address the
issue of scattering and tangling of concerns. Aspect oriented
programming provides support for cross-cutting concerns
(i.e. aspects) and for automatically propagating appropriate points of
execution in the code; nevertheless, this technology is not widely
established yet, and language support is still relatively limited in
C++. 

Two topics associated to concerns in Geant4 code are relevant to the
research areas considered in this project: the issue of endowing
objects - in particular, physics objects - of intrinsic verification
and validation capabilities (more in general, of analysis
capabilities), and dealing with secondary effects following a primary
interaction (e.g. the relaxation of an excited atom).
An exploratory study in these areas would be useful to evaluate the 
possibility of addressing concerns effectively in a simulation environment
like Geant4.

\section{Ongoing activities}

The activity currently in progress has an exploratory character: it evaluates 
various problem domains to identify the issues to be addressed, 
the requirements in the associated experimental domains and candidate 
technologies.

Regarding the main topic of research, the problem domain analysis has
identified the concept of ``mutability'' as a main issue in the context 
of transition between co-working condensed and discrete transport schemes.
The current research in software design explores the introduction of the
concept of ``mutants'' in the software design, and of ``stimuli'' capable
of triggering mutations. 
Related concepts, like reversible and spontaneous mutation, are subject to 
investigation too.

The introduction of the concept of mutability in physics-related objects
requires the identification of their stable and mutable states and behaviour,
and their fine-grained decomposition into parts capable of evolving, or 
remaining unchanged.
Two pilot projects are in progress to explore the capability of policy-based
class design to support this requirement in two large scale physics 
simulation domains:
a general-purpose one \cite{em_nss2009} and a ``track structure'' one.

Issues related to PIXE simulation are explored in another dedicated 
pilot project \cite{pixe_tns,pixe_nss2009}.

In parallel, a project focussed on the simulation of radioactive decay 
\cite{radioactive_nss2009}
explores issues related to the collaboration between electromagnetic and
hadronic components of the design associated with the prototypes explored
in the electromagnetic domain.

\section{Conclusion and outlook}
A R\&D project is in progress to address the capability of dealing with
multi-scale use cases in the same simulation environment based on Geant4: 
this requirement involves the capability of handling
physics processes according to different transport schemes. 
The first developments associated with
this project are described in greater detail in other contributions to 
the 2009 IEEE Nuclear Science Symposium Conference Record.
M. Kuster and S. Hauf acknowledege support by the Bundesministerium 
f\"ur Wirtschaft und
Technologie and the Deutsches Zentrum f\"ur Luft- und Raumfahrt - 
DLR under the grant number 50QR0902.

\section*{Acknowledgment}
The authors thank Sergio Bertolucci (CERN), 
Simone Giani (CERN), 
Vladimir Grichine (Lebedev Institute),
Bernd Grosswendt (formerly PTB, retired),
Alessandro Montanari (INFN Bologna),
Andreas Pfeiffer (CERN),
Reinhard Schulte (Loma Linda University),
Manju Sudhakar (ISRO and INFN Genova)
and Andrew Wroe (Loma Linda University)
for helpful discussions.

\end{document}